# Greenland Telescope Project

## --- Direct Confirmation of Black Hole with Sub-millimeter VLBI ---

Inoue, M.[§]; Algaba-Marcos, J.C.; Asada, K.; Chang, C.-C.; Chen, M.-T.; Han, J.; Hirashita, H.; Ho, P.T.P.; Hsieh, S.-N.; Huang, T.; Jiang, H.; Koch, P.M.; Kubo, D.Y.; Kuo, C.-Y.; Liu, B.; Martin-Cocher, P.; Matsushita, S.; Meyer-Zhao, Z.; Nakamura, M.; Nishioka, H.; Nystrom, G.; Pradel, N.; Pu, H.-Y.; Raffin, P.A.; Shen, H.-Y.; Snow, W.; Srinivasan, R.; and Wei, T.-S.

Academia Sinica Institute of Astronomy and Astrophysics

11F of Astronomy-Mathematics Building, AS/NTU, No.1, Sec. 4, Roosevelt Rd, Taipei 10617, Taiwan, R.O.C.

Blundell, R.; Burgos, R.; Grimes, P.; Keto, E.; Paine, S.; Patel, N.; and Sridharan, T.K.

Harvard-Smithsonian Center for Astrophysics

60 Garden Street, Cambridge, MA 02138, USA

Doeleman, S.S.; and Fish, V.

MIT Haystack Observatory

Off Route 40, Westford, MA 01886-1299, USA

and

Brisken, W.; and Napier, P.

National Radio Astronomy Observatory

Array Operation Center, P.O. Box O, 1003 Lopezville Road, Socorro, NM 87801-0387, USA

---

[§] Email address: inoue@asiaa.sinica.edu.tw

**Abstract**
A 12-m diameter radio telescope will be deployed to the Summit Station in Greenland to provide direct confirmation of a Super Massive Black Hole (SMBH) by observing its shadow image in the active galaxy M87. The telescope (Greenland Telescope: GLT) is to become one of the Very Long Baseline Interferometry (VLBI) stations at sub-millimeter (submm) regime, providing the longest baseline > 9,000 km to achieve an exceptional angular resolution of 20 µas at 350 GHz, which will enable us to resolve the shadow size of ~40 µas. The triangle with the longest baselines formed by the GLT, the Atacama Large Millimeter/submillimeter Array (ALMA) in Chile, and the Submillimeter Array (SMA) in Hawaii will play a key role for the M87 observations. We have been working on the image simulations based on realistic conditions for a better understanding of the possible observed images. In parallel, retrofitting of the telescope and the site developments are in progress. Based on three years of opacity monitoring at 225 GHz, our measurements indicate that the site is excellent for submm observations, comparable to the ALMA site. The GLT is also expected to make single-dish observations up to 1.5 THz.

## 1. Introduction

A direct observation demonstrating the existence of a Black Hole (BH) is one of the exciting goals in modern astronomy and physics. Experiments in the sub-millimeter (submm) Very Long Baseline Interferometry (VLBI) have shown the capability for the direct imaging of nearby BHs [*Doeleman et al.*, 2008, 2012; *Fish et al.*, 2013], and the image simulations have been made by several authors [e.g., *Lu et al.*, 2014]. The angular resolution achieved so far reaches the expected size of several tens of micro arcsec (µas) for some Super Massive BHs (SMBHs) with large apparent size. Submm observations also allow us to investigate the immediate vicinity of SMBHs, probing the circum-nuclear matter and its velocity field, both of which will be strongly affected by the strong gravity field.

However, the number of the submm VLBI telescopes is very limited (see Figure 5 and Table 1), and the baselines (*u-v* coverage) are not well distributed. Therefore, the development of new submm VLBI sites is a key to promote the SMBH science. Smithsonian Astrophysical Observatory (SAO) and Academia Sinica Institute of Astronomy and Astrophysics (ASIAA) were awarded a submm telescope that was constructed as a prototype for the Atacama Large Millimeter/submillimeter Array (ALMA). We are collaborating with the MIT Haystack Observatory and the National Radio Astronomy Observatory (NRAO). Accordingly, we have started the site survey for the telescope to establish a new submm VLBI station. In Section 2, we briefly describe the science targets using the telescope. In Section 3, the site selection process is explained as to how we selected the Greenland site, and hence the name Greenland Telescope (GLT). In Section 4, the expected submm VLBI network is shown from the GLT point of view, followed by a discussion of the current retrofitting status of the telescope in Section 5. In Section 6, the GLT system for VLBI operation is explained briefly, followed by a summary in Section 7.

## 2. Science Targets

The primary target is to image the shadow of the SMBH in M87 with the submm VLBI network using the GLT. Recently, Lu [2014] made several simulations with possible cases of telescopes. Furthermore, since the actual submm VLBI observation times will be short relative to the telescope availability, we plan to take advantage of the site by using the GLT for single-dish observations at frequencies up to 1.5 THz.

The image of a SMBH can be seen as a shadow against its bright accreting material and/or jets. As the Schwarzschild radius ($r_s$) is proportional to the SMBH mass, the apparent size of the shadow depends on the mass and distance. The diameter of the shadow is expected to be about 5 times of $r_s$ for a non-rotating BH, enlarged by a lensing effect of the strong gravity. The $r_s$ of Sgr A* is thought to be the angular diameter of about 10 µas from a mass estimate of $4.3 \times 10^6$ M$_\odot$ [*Gillessen et al.*, 2009], providing the largest apparent shadow diameter of 52 µas because of its close distance. The second largest one is in M87. Its large mass of $6.6 \times 10^9$ M$_\odot$ [*Gebhardt et al.*, 2011] gives $r_s \sim 8$ µas, showing an apparent shadow diameter of 42 µas comparable to Sgr A*, although the mass estimate is a little controversial [*Walsh et al.*, 2013]. Supported by imaging simulations [e.g., *Luminet*, 1979; *Falcke et al.*, 2000; *Dexter et al.*, 2012], we are now confident to be able to identify the shadows of SMBHs, Sgr A* and M87 being the best targets based on their apparent angular sizes. The shape of the shadow image depends on the black hole spin [e.g., *Falcke et al.*, 2000]. This in turn means that we could directly measure the key parameters of a SMBH, i.e., the mass and spin of a SMBH from its shadow image with submm VLBI, in addition to observing the other physical processes under the strong gravitational field, such as the accretion mechanisms onto SMBHs and the launching mechanisms of relativistic jets.

Although Sgr A* is likely to have the largest apparent size of the black hole shadow, it shows rapid intensity variations within a day even at mm and submm wavelengths [*Miyazaki et al.*, 2004; *Fish et al.*, 2011]. This is probably due to the small mass of the black hole. Such rapid variations make it difficult to generate non-smeared synthesized images of Sgr A* with VLBI. On the other hand, the mass and size of M87 are $10^3$ times larger than that of Sgr A*, and it is likely to have variations on much longer time scales. The SMBH mass of M87 suggests a typical orbital timescale of 3-35 days depending on the SMBH spin. Therefore, unlike Sgr A*, the structure of M87 may not change during an entire day of VLBI image synthesis. On this point, M87 is a better candidate than Sgr A* for the VLBI imaging synthesis using the Earth rotation. Moreover, it would be possible to produce a sequence of images of dynamical events in M87, occurring over many days. M87 is, thus, a promising target to image a SMBH shadow with its bright accreting material around it. Furthermore, as shown in Figure 1, the tracing back of the jet radius toward the core is one of the crucial tests to identify the launching point of the jet. Consequently, this provides an important hint on the jet formation mechanism. If the parabolic shape of the jet extends towards the innermost region, i.e., a distance of 2 $r_s$ from the SMBH, it is likely that the jet is originated within the Innermost Stable Circular Orbit (ISCO) of the accretion disk around the SMBH or the SMBH itself. Submm VLBI observations of the jet radius make it possible to reveal the launching point and mechanism of the relativistic jet [*Asada and Nakamura,* 2012; *Nakamura and Asada*, 2013b; *Asada et al.*, 2014].

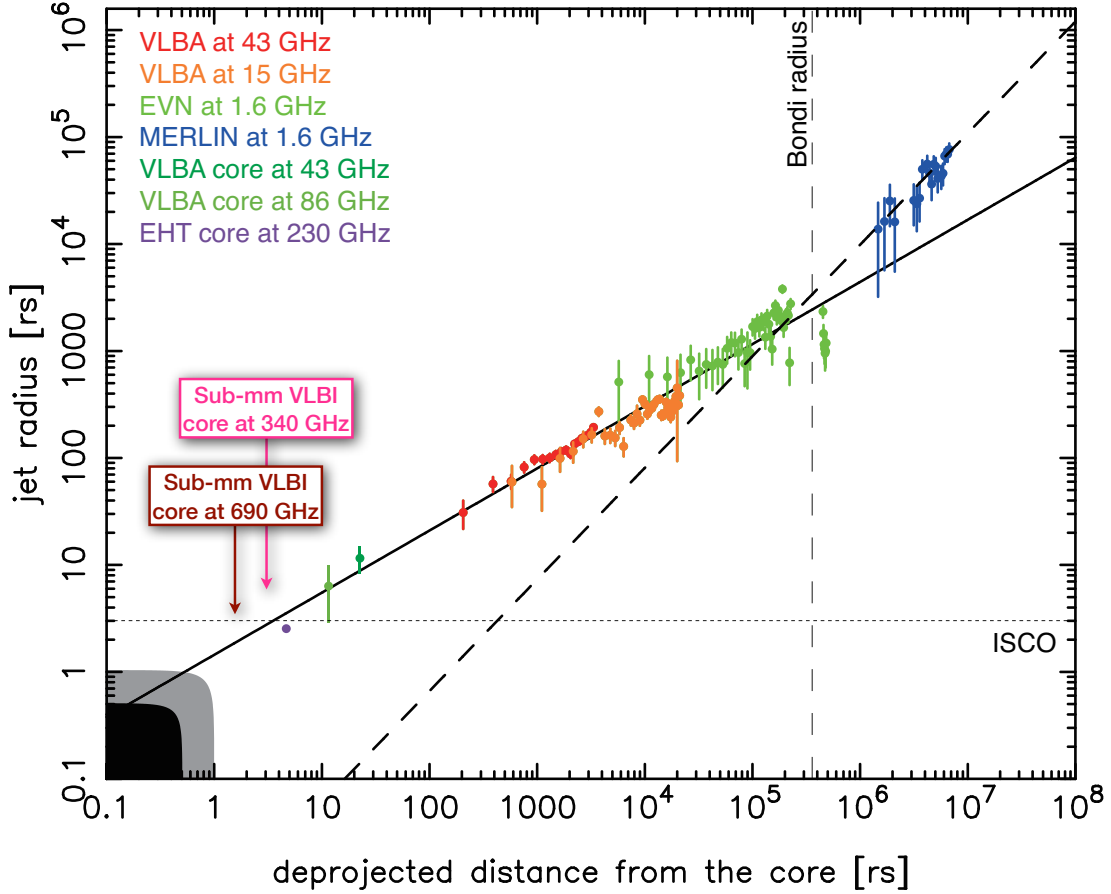

Figure 1. The jet radius of M87 against the distance from the core in unit of Schwarzschild radius $r_s$. Close to the core, the submm observation of jet radius could suggest the launching point and its mechanism [*Asada and Nakamura*, 2012; *Nakamura and Asada*, 2013b]. This plot also shows the shape of the jet. On the upstream side from the core to several times of $10^5$ $r_s$, the shape is parabolic (solid line), and then changed to conical (dashed line), suggesting the magnetohydrodynamic mechanism at work in both acceleration and collimation of jet in the upstream region [*Asada and Nakamura*, 2012; *Nakamura and Asada*, 2013a]. Based on the VLBI core shift measurement [*Hada et al.*, 2011], the expected positions of submm VLBI core at 340 and 690 GHz are indicated.

Although M87 provides a good opportunity to study the jet mechanism, this may introduce complexity to investigate the shadow image. Image simulations based on realistic models of accretion flows and jets will be a key to understand the observed image correctly. From this point, we have been studying some image simulations shown in Figure 2. This model shows a case of only a black hole without jet contribution, for simplicity. High-resolution and high-sensitivity observations using the phased ALMA are discussed and reported by Fish et al. [2013], a submm VLBI group including us.

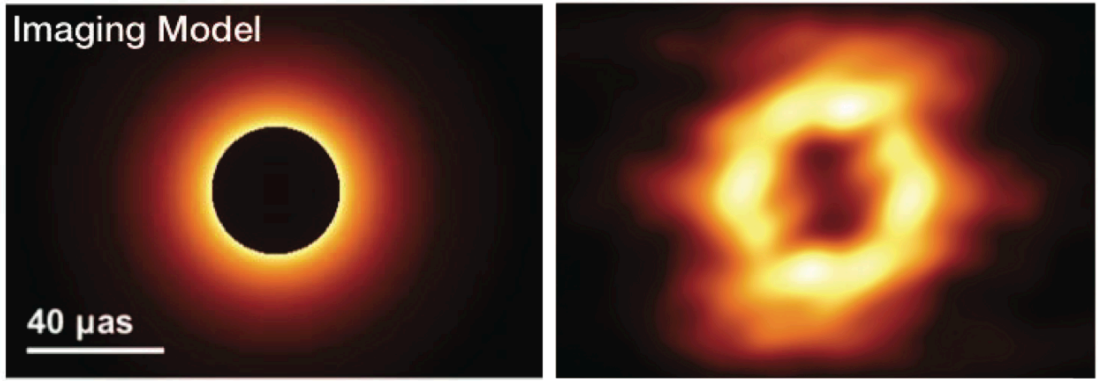

Figure 2. Image simulation of the SMBH shadow of M87. (left) The model is made using ray-tracing, assuming a non-rotating, 6 billion solar mass SMBH surrounded by optically thin, free falling material. For simplicity, the model does not include the jet contribution. (right) Based on the model image, the observed image is simulated for a submm VLBI observation including the GLT at 345 GHz. In the visibility data, expected thermal noise is added, and CLEAN process is applied.

As described in Section 3, the telescope site in Greenland has very low water vapor content and is, therefore, good for submm single-dish observations. On this point, the science cases of the single-dish observations are somewhat similar to those of the Cerro Chajnantor Atacama Telescope (CCAT), and the GLT is expected to play a role as the forerunner of CCAT with a frequency coverage up to 1.3-1.5 THz. A major advantage of moving to higher frequencies into the THz regime is that thermal dust continuum emission will be measured around its peak in the spectral energy distribution for nearby objects whose cosmological redshift is negligibly small. Moreover, the GLT angular resolution (4″ at 1.5 THz) enables us to spatially resolve the individual star formation sites within nearby molecular clouds (< 1 kpc). These two points would be big advantages especially in searching for less massive pre- and proto-stellar populations (e.g., brown dwarf mass or even less massive sources) and constrain the core mass function at the lower mass end. Furthermore, in combination with multi-frequency data, this will allow us to measure spectral energy distributions of pre- and proto-stellar sources and determine their cool-gas temperatures of ∼10 K, thus providing access to the earliest evolutionary phases of star formation. For nearby galaxies, spatially resolved studies of dust emission are possible, enabling us to trace the star formation activities in a spatially resolved way. The high angular resolution also overcomes the confusion-limited surveys of high-redshift galaxies in the past at far-infrared wavelengths.

There are also some unique atomic and molecular lines at THz frequencies. Combined with the GLT's high angular resolution and spectroscopic capability, it is possible to create three-dimensional maps (R.A., Dec., and velocity) of THz lines: THz molecular lines (e.g., CO, HCN) will probe warm (300–500 K) molecular regions in the vicinities (< 10 AU) of forming protostars, tracing inner expansions driven by radiation pressures and outflows from central protostars. Combined with cold gas inflow from outer envelopes traced by lower-J lines, it is possible to study the stellar mass growth in the proto-stellar cores. Additional lines accessible in the THz windows will be groups of atomic fine-structure lines (e.g., [N II]), which trace diffuse transitional regions from ionized or atomic gas to molecular gas in the interstellar medium. Combined with the kinematic information, cloud formation mechanism, which can be considered as the first step of star and planet formation, or the disruption mechanism of clouds by the formed stars, can be studied. THz lines of hydrides (e.g., CH, $OH^+$, $SH^+$) are also interesting molecular lines, since these molecules are one of the first molecules to form in the chemical processes, and therefore important to study interstellar chemistry.

**3. Site Selection and Construction Plan**

Site survey for the submm VLBI station was made under the two main criteria: (1) to have excellent atmospheric condition at higher frequencies, and (2) to form long baselines and more complete *u-v* coverages with existing and planned submm VLBI sites, particularly ALMA in Chile and the SMA in Hawaii. The phased ALMA system will be a high-sensitivity VLBI station. Through our involvement in the SMA operation and the ALMA phase-up project (APP) we are closely linked to and familiar with both interferometers, as mentioned in the next section. Satellite data of Precipitable Water Vapor (PWV) provided some low PWV areas on the Earth, and most of these areas already have some radio telescope facilities, like ALMA, the South Pole Telescope, etc. Among these areas, Tibet and Greenland are left untouched. Tibet does not have common visibility with ALMA, while Greenland has long baselines with both ALMA and the SMA for M87 (see Figure 3). In addition, there are already some research facilities at the Summit Station on the ice sheet, supported by NSF, USA, located at 72.57°N and 38.46°W.

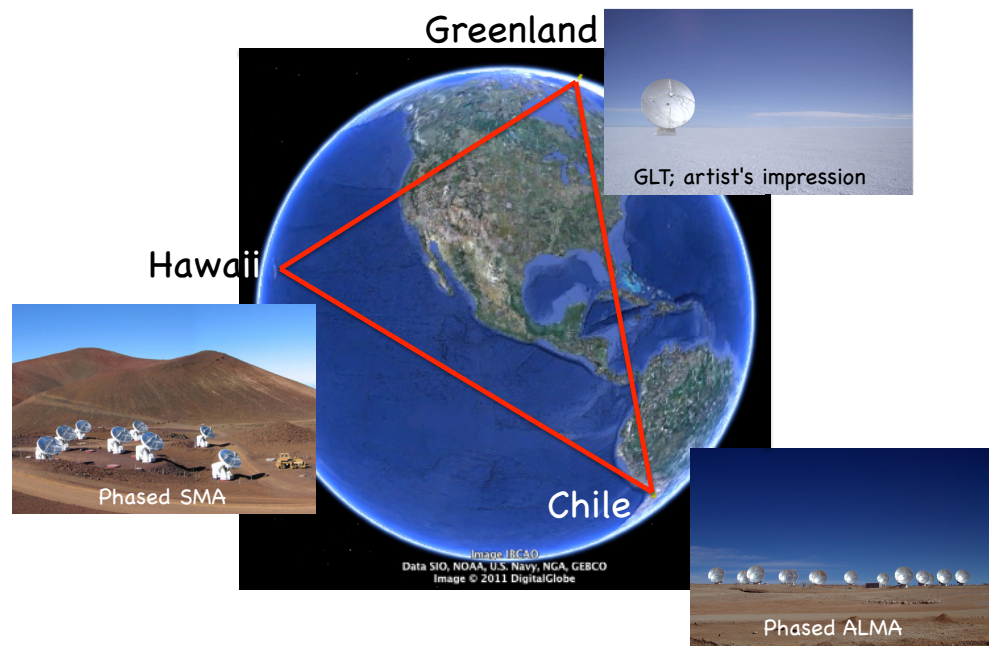

Figure 3. A big VLBI triangle constituted by the GLT, ALMA, and the SMA. The latter two are closely related to the GLT project (see the text). The longest baseline is longer than 9,000 km, providing an angular resolution of 20 μas at 350 GHz. For M87, the shadow size is expected to be around 40 μas, and the shadow image is highly expected with this submm VLBI network, collaborating with the other submm

VLBI stations (see Figure 2).

The altitude of the Summit Station is 3,200 m above sea level with the lowest temperature of approximately -70 ˚C. Although the altitude is not so high compared to the other submm sites like ALMA and SMA, the low temperature will provide a good dry atmospheric condition for submm observations.

We have deployed a tipping radiometer at 225 GHz in the summer 2011 at the Summit Station to monitor the opacity. *Matsushita et al*. [2013] discussed the monitoring results for two years, showing comparative quality of opacity to that of the ALMA site. Figure 4 shows the statistics during winter seasons from Oct. 2011 to Mar. 2014. The best opacity at 225 GHz was 0.021, and the 25% quartile, median, and 75% quartile were 0.048, 0.063, and 0.085, respectively with the most frequent opacity of 0.04. These statics are compatible with those at the ALMA site, and in summer, it is better than that of ALMA [*Matsushita et al*., 2013]. These statistics are very promising for the submm VLBI and single-dish observations, although the effect of phase fluctuation needs to be taken into account for VLBI observations. As Greenland does not have high mountains, air turbulence may not be serious. An inversion layer at around 10 m from the snow surface may be formed when wind is weak. The phase fluctuation by this should be negligible, but the antenna structure could suffer from the temperature gradient as indicated in Section 5.

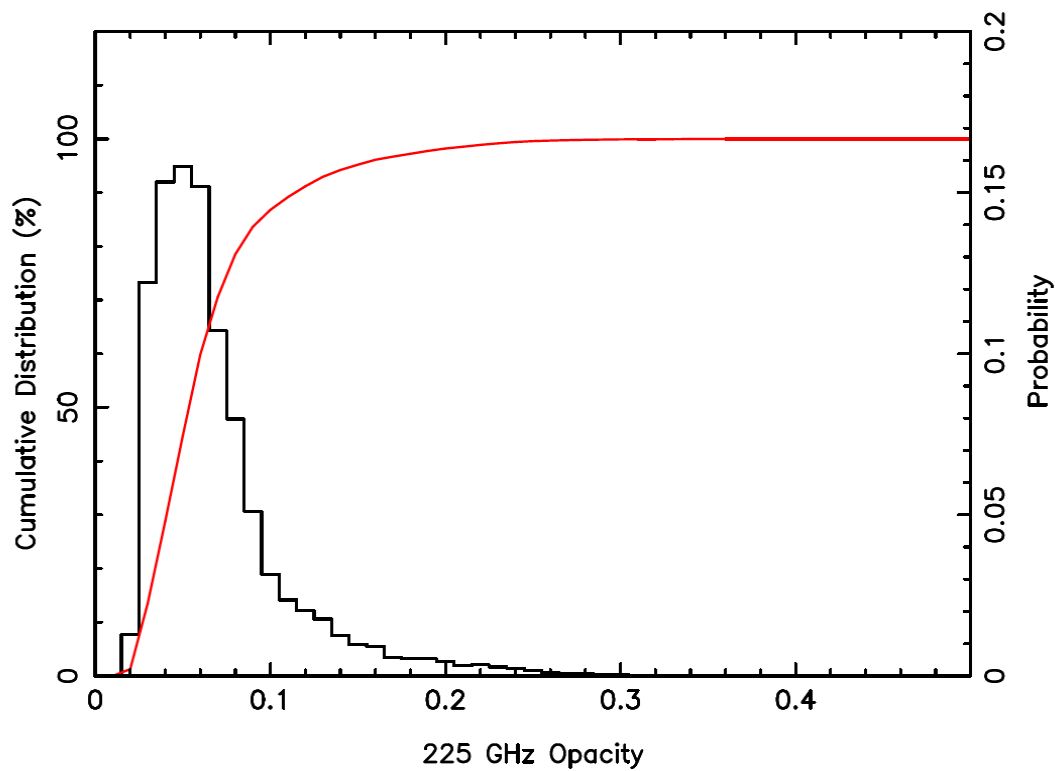

Figure 4. Statistics of the opacity monitoring at the Summit Station in Greenland, measured by our tipping radiometer at 225 GHz. The statistics shows only the winter season: 1 Oct. 2011-31 Mar. 2012, 1 Oct. 2012-31 Mar. 2013, and 1 Oct. 2013-14 Mar. 2014. The left scale shows the cumulative distribution indicated by the red curve, and the right scale is for the histogram of 60,517 measurements. The best opacity was 0.021, and the 25% quartile, median, and 75% quartile were 0.048, 0.063, and 0.085, respectively with the most frequent opacity of 0.04.

We have been working on the construction of the telescope foundation and infrastructure, collaborating with the CH2M HILL Polar Service (URL: http://cpspolar.com/project-locations/greenland/). A new power station will be constructed to provide electrical power for the telescope and associated facility.

**4. Submm VLBI Network**

As seen in Figure 3, the baselines between Greenland, Hawaii, and Atacama in Chile constitute a big triangle for M87, extending more than 9,000 km long, to achieve an angular resolution of 20 µas at 350 GHz. Table 1 shows baseline lengths (D) and angular resolutions ($\lambda$/D) in µas at 230 GHz together with other submm telescopes: IRAM 30-m (Pico Veleta, Spain), SMTO (Arizona, USA), LMT (Mexico), CARMA (California, USA). These telescopes will provide a good *u-v* coverage, as shown in Figure 5, to make a plausible quality image of the M87 shadow. Many of these telescopes will also constitute a VLBI network at 350 GHz, providing high angular resolution images [see *Lu et al.*, 2014]. The GLT at the Summit Station has good baselines between both European and US submm telescopes. The North-South baselines subtended by the GLT and other telescopes also provide a good opportunity to investigate the structure of the inner jet in M87 which is oriented in East-West direction.

Table 1. Baseline length and spatial resolution

|  | PdBI | IRAM | ALMA | CARMA | SMTO | LMT | SMA | GLT |
|---|---|---|---|---|---|---|---|---|
| PdBI (B) | - | 234 | 29 | 32 | 32 | 31 | 25.2 | 70.7 |
| IRAM (V) | 1,147 | - | 31.2 | 31.3 | 32.0 | 32.2 | 24.7 | 62.2 |
| ALMA (A) | 9,400 | 8,624 | - | 34.0 | 37.5 | 49.2 | 28.5 | 28.1 |
| CARMA (C) | 8,540 | 8,586 | 7,906 | - | 289.7 | 93.9 | 67.1 | 48.3 |
| SMTO (S) | 8,481 | 8,402 | 7,176 | 928 | - | 136.9 | 58.1 | 46.8 |
| LMT (L) | 8,686 | 8,352 | 5,469 | 2,863 | 1,964 | - | 45.9 | 40.4 |
| SMA (H) | 10,671 | 10,907 | 9,448 | 4,009 | 4,627 | 5,858 | - | 33.2 |
| GLT | 3,805 | 4,323 | 9,572 | 5,570 | 5,742 | 6,647 | 8,096 | - |

Baseline length D (km) is shown in lower left half, and spatial resolution in upper right half in $\lambda$/D (µas) at 230 GHz. A character in parentheses is a symbol of station in Figure 5.

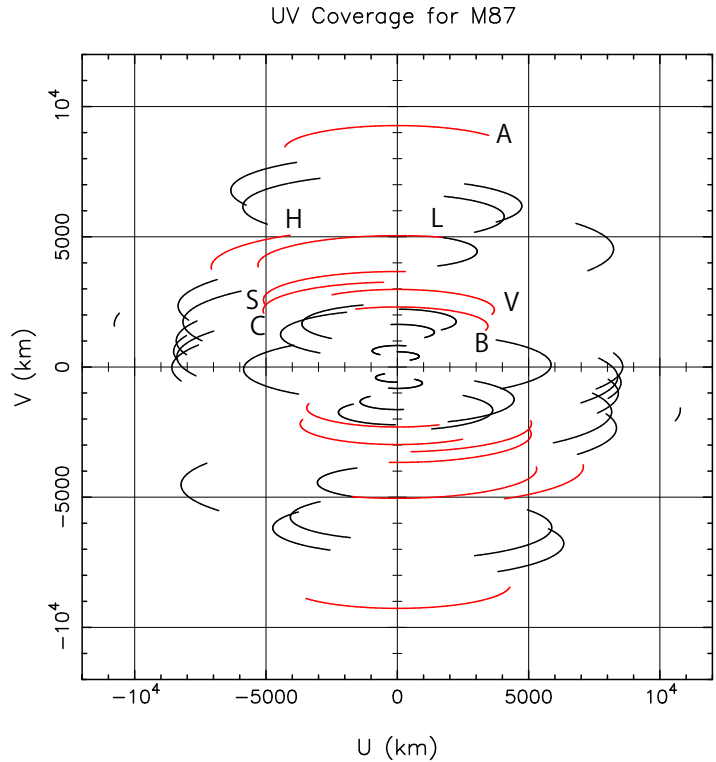

Figure 5. The expected *u-v* coverage towards M87 with the GLT and other existing and planned submm telescopes: GLT, ALMA (A) in Chile, SMA (H) in Hawaii, SMTO (S) in Arizona, CARMA (C) in California, LMT (L) in Mexico, IRAM 30-m telescope (V) at Pico Veleta, Spain, and IRAM Plateau de Bure Interferometer (B) in France. The red lines are the GLT contribution, and characters beside them indicate the corresponding stations indicated in the parentheses. The observation time with the SMA (H) is the shortest, about 1 hour with an elevation angle $\geq 20°$, and 4 hours with their elevation limits. In addition to the longest baseline in North-South direction, the GLT baselines are filling in 3,000-5,000 km regions. In Table 1, the baseline lengths (D) with their corresponding angular resolution ($\lambda/D$) are given.

ASIAA and SAO are institutions that have close relationships to both the SMA and ALMA. The project to phase-up all of the ALMA antennas is in progress with international collaboration. Similarly, the phase-up of the SMA is already at work led by SAO and Haystack Observatory. Once completed, this triangle is expected to provide high sensitivity and high angular resolution at 350 GHz and even higher frequencies. At 230 GHz, many other submm telescopes will collaborate with this big triangle to form a good submm VLBI network. Lu et al. [2014] provide the System Equivalent Flux Density (SEFD) of the various telescope combinations.

## 5. Retrofitting the Telescope

After being awarded and taking possession of the ALMA prototype telescope in April 2011, we conducted initial performance tests as the telescope was left unattended since the original characterization [*Magnum et al.*, 2006]. The photogrammetric measurements of the main surface showed a good surface accuracy of about 52 µm rms, after one iteration of adjustment. The pointing accuracy was 2 arcsec rms with a new wide-field-of-view optical guide telescope. These values were near the measurement limits under the tight test constraints, and we concluded that the telescope was kept in good condition without any serious damages and degradations. Thereafter, we started discussion on retrofitting the telescope for the cold environment on Greenland with its original vender Vertex Antennentechnik GmbH.
In terms of requirements we generally follow the ALMA telescope specifications. The main differences from the ALMA specifications are as follows:

(1) Operating conditions
  Ambient temperature        -50 ºC < T < 0 ºC
  Vertical temperature gradient     +12 K > $\Delta$T > -1 K
  Ambient air pressure        671 mbar ± 12 mbar
  Wind             11 m/s average wind speed
(2) A steel spaceframe tower interfacing the telescope to the foundation.
(3) A nutator in the subreflector system, compensating for smaller maximum slew speed and acceleration in azimuth and elevation drives that do not allow fast switching.
(4) A tertiary mirror system to accommodate VLBI receivers, multi-pixel camera, and multi-beam receiver in the receiver cabin.
(5) Enclosures attached for receiver compressors, elevation gears, and other devices.
(6) Heating pads on the backside of the surface panels for de-icing.

The vertical temperature gradient is defined throughout the antenna height, due to a possible inversion layer at several tens of meters height relative to the foundation. The gradient $\Delta$T = +12 K means the top of antenna is 12 K higher than the bottom, or foundation. The spaceframe tower prevents snow from accumulating around the telescope. As the Summit Station is on an ice sheet of about 3,000-m thick, the telescope foundation will be formed from packed snow. Because of the snow foundation, rigidness of the spaceframe, and degradation of the steel drive gears due to low temperature, we will not perform fast switching of the main reflector. The item (3) above was added to compensate for the removal of fast switching. .
On the backside of the surface panels, heating pads are attached to prevent ice or frost from forming on the antenna surface. When ice or frost formation is forecasted, the temperature of the panels will be set several Kelvin higher than the ambient temperature. The enclosures are designed to cover the elevation driving motors, cryogenic compressors for the receivers, as well as other items. The receivers are brought into the receiver cabin through an enclosure that provides a workspace and serves as buffer between the outdoors and the cabin interior. The azimuth and elevation bearings and electrical cables are modified to match the cold environment.
Figure 6 shows the present concept of the receiver cabin. The tertiary mirror near the Cassegrain focus leads the beam to the receivers located in the upper right or lower left side. By removing the tertiary mirror, the beam goes down to a higher frequency instrument located at the central area. To ensure the higher frequency observations, the antenna primary surface will be set to high accuracy using holography to enable efficient operation at up to THz bands. The required systems, 90-GHz band room-temperature receivers and beacon have been mostly built and tested.

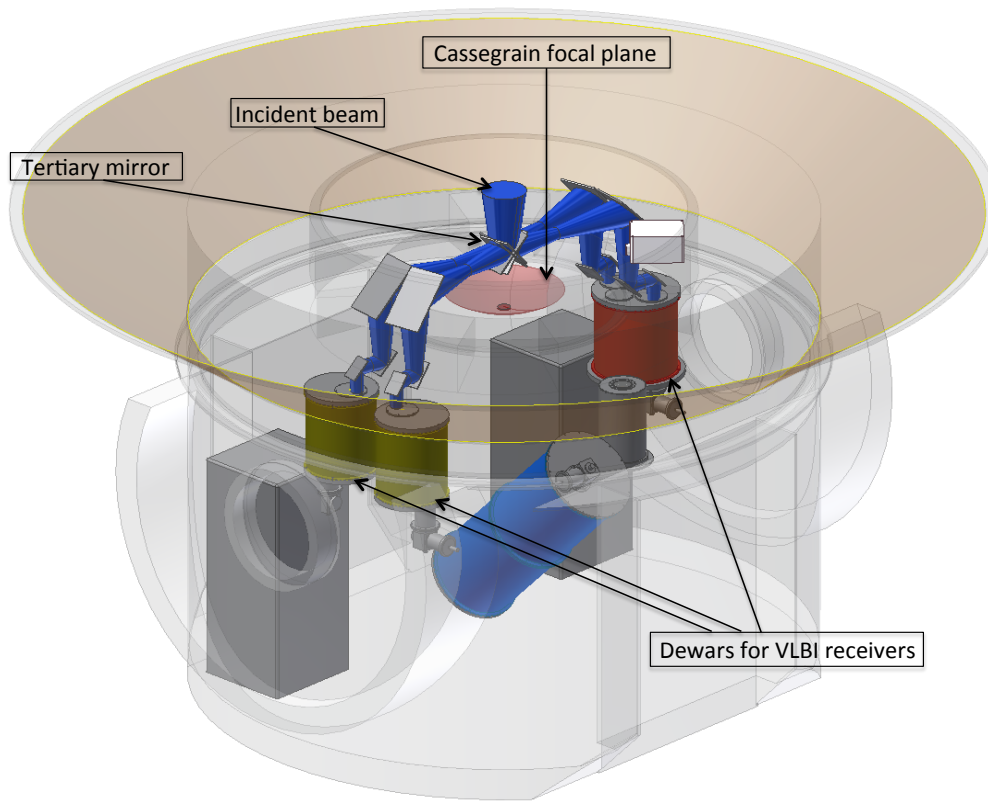

Figure 6. Conceptual drawing of the beam optics in the receiver cabin. Above the Cassegrain focus, a tertiary mirror reflects the beam to upper right and lower left sides where dewars of VLBI receivers are set. All the beams are transferred into the dewars keeping the same beam parameters with those of the Cassegrain focus. The curved surface below the tertiary mirror shows the Cassegrain focal plane, which is used for the high frequency cameras and multi-beam horn receivers.

For the submm VLBI observations, receivers at 230, 350, and 650 GHz with dual circular polarization are planned and are under construction. The beam is directed to a selected VLBI receiver using the secondary and tertiary mirrors (see Figure 6). In addition, a low-frequency receiver at 86 GHz will be installed for initial VLBI system checkout and fringe search during the commissioning phase. This receiver will also be used for the antenna pointing check using SiO maser sources. Heterodyne multi-feed receivers and multi-pixel bolometers up to 1.5 THz have been discussed. The GLT is also expected to be a test bench of receiver developments at higher frequencies. The weather condition in wintertime is comparable to that of the ALMA site and even the South Pole in the local wintertime.

All telescope components will be reassembled and checked out once before setting up at the Summit Station. Transportation to access the Summit Station is once a week in the summer and every three months in winter, and therefore requires a well planned and coordinated sequence of deliveries. Regarding the transportation of the VLBI observing data, we will not make VLBI observations in midwinter. During the midwinter in the Northern Hemisphere, the weather condition at ALMA is not good, and many other submm telescopes in the Northern Hemisphere are not easy to obtain VLBI observation time. We will coordinate the VLBI observations in the spring and autumn, when more frequent transportation of VLBI data is available. We will store key spare components at the site because the turnaround time during the winter can be quite long. We are planning to ship the telescope components to Thule Air Base located on the northwest cost of Greenland in 2015, followed by assembly and first light in Thule during 2016. Sometime in 2018/19, the GLT will be established at the Summit Station.

**6. VLBI System**

A hydrogen maser will be used as the master local oscillator system and has already been purchased and delivered. Two separate GPS systems are planned, one to provide accurate timing for VLBI and monitoring the drift of the hydrogen maser, and the other to provide accurate telescope position coordinates. The Summit Station is located on a thick ice sheet and is thought to move roughly 1 m per year. The telescope position should be measured for each VLBI observation with an accuracy ≤ 1 cm.

The receivers at 86, 230, and 350 GHz are mainly used for VLBI observations. These have the same cartridge type as the ALMA receivers. As the VLBI receivers are almost independent of the single-dish receivers, we could easily switch the observation mode between VLBI and single-dish operations.

The VLBI group in ASIAA has been participating in the APP led by MIT Haystack Observatory [URL: https://deki.mpifr-bonn.mpg.de/ALMA_Phasing_Project/Public_Page_-_ALMA_Phasing_Project].

Although the APP is not directly related to the GLT, the high sensitivity of the phased-up ALMA will largely contribute to the longer baselines formed by the GLT.

One of ASIAA's tasks in the APP is to solve the issue of different sampling rates between ALMA and the other traditional VLBI stations. We will solve this issue by changing the frequency band of the VLBI stations in the frequency domain, instead of arranging the sampling rate in the time domain. This has been implemented in the Distributed FX-style software correlator (DiFX) system [*Deller et al.*, 2011] as an enhancement of DiFX. For this implementation, ASIAA purchased a CPU cluster to run DiFX and its enhancement system. With this DiFX correlator, we are planning to correlate the commissioning data of the GLT. Based on this knowledge, we are discussing to have a DiFX correlator dedicated to the submm VLBI observations. The correlator is requested to process a huge amount of data up to 64 Gbps per station, and may need to collaborate with other institutes for the operation. This correlator will play an important role as one of the submm VLBI correlator sites.

**7. Summary**

ASIAA and SAO were awarded an ALMA prototype antenna, collaborating with MIT Haystack Observatory and NRAO. We identified an excellent site in Greenland to set up this antenna as a submm VLBI station. Based on this site location, we named this telescope and project as the Greenland Telescope (GLT) Project. The main science driver is to produce a shadow image of the SMBH in M87 to obtain direct proof of the black hole, using the other submm systems like the SMA in Hawaii, ALMA in Chile, etc. With the GLT-SMA-ALMA triangle, an angular resolution of 20 μas will be achieved at 350 GHz. With such high angular resolutions, we will be able to identify the shadow image of M87. In addition, the GLT provides long North-South baselines which will be beneficial for high angular resolution observations of the jet in M87. In parallel, several science cases have been discussed for the single-dish observations. As the PWV contents are low, the GLT is expected to operate up to 1.5 THz.

The retrofit of the telescope for the cold environment in Greenland is in progress, together with the arrangements of the infrastructure at the Summit Station. For submm VLBI, receivers at 230 and 350 GHz with dual polarization are under construction. For the fringe finding and pointing check, a low-frequency receiver at 86 GHz is being built. A new beam optics is designed to separate VLBI and single-dish receivers.

Our VLBI group has joined in the APP to increase the sensitivity of the submm VLBI observations. For the GLT VLBI observations, ALMA is a key station to get higher angular resolution. The high sensitivity also helps a lot, particularly for the longer baselines. Throughout the APP experience, we are discussing to have a dedicated correlator for submm VLBI.

**Acknowledgements**
The Greenland Telescope (GLT) Project is a collaborative project between Academia Sinica Institute of Astronomy and Astrophysics, Smithsonian Astrophysical Observatory, MIT Haystack Observatory, and National Radio Astronomy Observatory. Figure 1 is from Nakamura and Asada [2013a], with kind permission of the European Physical Journal (EPJ). We thank the Integrated Characterization of Energy, Clouds, Atmospheric state, and Precipitation at Summit (ICECAPS) project for valuable supports of a place to stand for our radiometer in their Mobile Science Facility (MSF). The ICECAPS project also provided us their radiosonde data, surface meteorological data, and radiometer data, to have a good picture of the site condition. We would also like to thank the National Science Foundation (NSF) for operating and funding Summit Station, the National Oceanographic and Atmospheric Administration (NOAA) for allowing access to their weather data, the Air National Guard 109th airlift wing for providing transportation, and a special thank to Konrad Steffen, at the Cooperative Institute for Research in Environmental Sciences (CIRES) for graciously sharing precious environmental data. The data in Figure 4 are available upon request to the corresponding author (Inoue, M.).